 \documentstyle[11pt,aaspp]{article}
\newcommand{\BE}{\begin{equation}}
\newcommand{\BEAL}{\begin{eqnarray}}
\newcommand{\EE}{\end{equation}}
\newcommand{\EEAL}{\end{eqnarray}}
\def\_#1{_{\scriptscriptstyle #1}}
\def\&#1{^{\scriptscriptstyle #1}}
\def\vsm{\langle v^2 \rangle}
\def\kms{kms\&{-1}}
\def\cmss{cm~s\&{-2}}
\def\lsun{L\_{\odot}}
\def\msun{M\_{\odot}}
\def\mlsunl{(M/L)\_{\odot}}
\def\mlsun{\left({M\over L}\right)\_{\odot}}
\def\fk{f\_k(\vec r,\vec v)}
\def\rs{r_{s}}
\def\rh{R\_{1/2}}
\def\ao{a_{o}}
\def\ho{H_0}
\def\grad{\vec \nabla}
\def\rar{\rightarrow}

\def\deriv#1#2{{d#1\over d#2}}

\def\l{\lambda}

\def\o{\omega}

\def\d{\delta}
\def\a{\alpha}
\def\b{\beta}
\def\c{\gamma}
\def\m{\mu}
\def\e{\eta}
\def\eh{\hat\eta}
\def\z{\zeta}
\def\zh{\hat\zeta}
\def\r{\rho}

\def\ff{\varphi}
\def\z{\zeta}
\def\t{\tau}
\def\mo{\m_0}
\def\ro{r_0}
\def\rhoo{\r_0}

\def\bk{\par\noindent}

\def\s{\sigma}
\def\sr{\sigma_r}
\def\st{\sigma_t}
\def\sp{\sigma_\perp}
\def\mlm{\left({M\over L}\right)\_{M}}

\def\mlnl{(M/L)\_{N}}
\def\div{\vec\nabla\cdot}
\def\ofpg{{1 \over 4\pi G}}
\def\vr{\vec r}
\def\vv{\vec v}
\def\gf{\grad\ff}
\def\dtr{d^2r}
\def\dtv{d^2v}
\def\hm{\hat\mu}
\def\hb{\hat\b}
\def\emo{m_0}
\begin{document}
\title{Large-scale filaments--Newtonian vs. modified dynamics}
\author{ Mordehai Milgrom}
\affil{Department of Condensed-Matter
 Physics, Weizmann Institute of Science  76100 Rehovot, Israel}
\begin{abstract}
\cite{elt96} (ELT) have recently proposed a method
for estimating the dynamical masses of large-scale filaments, whereby
the filament is modeled by an infinite, axisymmetric, isothermal,
 self-gravitating, radially virialized cylinder, for which ELT derive
a global relation between the (constant) velocity dispersion and the
total line density.
We first show that the model assumptions of ELT can be relaxed
materially: an exact relation between
the root-mean-square velocity and the line density can be derived for
 any infinite cylinder (not necessarily axisymmetric), with
an arbitrary constituent distribution function (so isothermality need
 not be assumed).
We then consider the same problem in the context of the modified
dynamics (MOND). After a brief comparison between the scaling
properties in the two theories,
we study two idealized MOND model filaments, one with assumptions similar
to those of ELT, which we can only solve numerically, and another,
which we solve in closed form.
A preliminary application to the same segment of
 the Perseus-Pisces filament treated by ELT, gives
MOND $M/L$ estimates of order $10\mlsunl$, compared with
the Newtonian value
 $M/L\sim 450(\ho/100 \kms Mpc^{-1})\mlsunl$, which ELT find.
 In spite of the large uncertainties still besetting the analysis,
this instance of MOND application is of particular interest because:
1. Objects of this geometry have not been dealt with before. 2. It
pertains to large-scale structure. 3. The typical accelerations
 involved are the lowest so far encountered in a semi-virialized
 system--only a few percents of the critical MOND acceleration--leading
to a large predicted mass discrepancy.
\end{abstract}
\section{Introduction}
\par
In a recent paper \cite{elt96} (ELT) describe a method
for estimating the dynamical masses of large-scale filaments.
Elementary considerations lead one to expect that when such a filament
is virialized in the radial direction its total line density
 (mass-per-unit-length), $\mo$, and its typical velocity dispersion,
$\s$, are related by $G\mo\sim\s^2$. For concreteness, ELT  consider
a model of these filaments consisting of an infinitely long,
 axially symmetric, self-gravitating cylinder whose constituent
velocity dispersions are constant (we call it isothermal, even if
the velocity distribution need not be ``thermal'').
 For such cylinders, ELT derive the exact relation
 \BE G\mo=2\sp^2,  \label{iiii} \EE
where $\sp$ is the velocity dispersion along a line of sight
 perpendicular to the cylinder axis, averaged over the whole cylinder.
ELT study carefully the applicability of this relation by testing
its performance on
N-body-simulated filaments. Finally, they apply this relation to a
segment of the Perseus-Pisce filament and estimate for it a B-band
$M/L$ ratio of about $450\mlsunl$, implying large quantities of
 dark matter.
\par
 The modified dynamics (MOND) was proposed as an alternative theory
to Newtonian dynamics with a view to eliminating the need for dark
matter [\cite{mil83}]. It posits that in the limit where
the typical accelerations in a system governed by gravity are much
smaller than some value ($\ao\sim  10^{-8}\cmss$)--the
 deep-MOND limit--the mass, $M$, radius, $r$,
 and acceleration, $a$, are related by $a^2/\ao\approx MG/r^2$,
and not by the standard $a\approx MG/r^2$. So, the smaller the
mean acceleration in a system, the larger the expected
mass discrepancy deduced by a Newtonian analysis.
\par
MOND has been tested extensively on galaxies
 [\cite{bbs91}, \cite{san96}], and it is natural to check how well it
does in the context of large-scale structure.
As explained in \cite{mil89},
MOND is not developed enough to afford application to most large-scale
structures, with over-density not much exceeding unity, or where the
Hubble flow is important.
Because large-scale filaments seem to be at least radially virialized,
they offer a unique opportunity to apply MOND to large-scale structure.
Inasmuch as a filament is not completely virialized, and extends to
radii where the Hubble flow is important, our analysis can be valid
only at smaller radii.
\par
 Large-scale structures constitute an extreme from the
point of view of MOND because they involve accelerations that
are the lowest in the range so far observed, and they
 carry the usual importance and interest of extremes:
Values as low as $0.1\ao$ are measured at the outskirts of dwarf spirals
[e.g. \cite{san96}], while, as we shall see below,
 in the Perseus-Pisces filament $a$ is only a few percents of
$\ao$.
\par
Section 2 brings some generalities concerning cylindrical and spherical
structures in Newtonian dynamics and in MOND, and extends the
ELT result to more general cylinders.
In section 3 we describe two models for self-gravitating
 infinite cylinders in MOND.
In section 4 we consider the Perseus-Pisces filament.

\section{Virial relations for cylindrical and spherical
 systems in Newtonian dynamics and in MOND}
\par
 Virialized spherical systems of size $r$ and velocity dispersion
$\s$, have a mean acceleration, $\sim \s^2/r$, that
 roughly equals $MG/r^2$
in Newtonian dynamics, and $(MG\ao/r^2)^{1/2}$ in MOND
(very-low-acceleration limit). Thus, in the former the virial relation
$M\sim \s^2 r/G$ contains the system's radius, while in the latter the
analogous relation $M\sim \s^4/G\ao$ does not. For cylinders, the
tables are turned, and since the Newtonian mean acceleration is
$\sim \m G/r$, while in MOND it is $\sim(\m G\ao/r)^{1/2}$, we have in
Newtonian dynamics $\m G\sim\s^2$, while in MOND we have
$\m G\sim \s^4/r\ao$.
\par
 We see then that
the Newtonian cylindrical case is akin to the spherical, deep-MOND
case, the crucial common property
 being the logarithmic behavior of the
potential at infinity. In fact,
 the isothermal axisymmetric cylinder discussed by ELT, and
 the deep-MOND, isothermal spheres described in \cite{mil84} are
special cases, with dimensions two, and three, respectively,
of $D$-dimensional, self-gravitating, isothermal spheres, held together
by a potential, $\ff$, that satisfies a Poisson-like equation
\BE \div[\vert\gf\vert^{\a}\gf]=A\r, \label{karte} \EE
where $\a=D-2$ is chosen such that the potential of an isolated mass
is logarithmic at infinity (for $D=2$ we get the usual Poisson equation,
for $D=3$ we get deep MOND).
The general density law of such spheres is (\cite{mil96})
\BE \r(r)\propto {(r/\ro)^{-(D-1)\b}\over \left[1+(r/\ro)^{{D\over D-1}
-\b}\right]^D}  \label{lautr} \EE
($\b$ is the anisotropy parameter--see the next section).
These spheres satisfy a mass-velocity-dispersion relation of the form
\BE \vsm^{D-1}=MA\a\_{D}^{-1}\left({D-1\over D}\right)
^{D-1},  \label{dusna} \EE
where $\vsm$ is the mass-weighted root-mean-square velocity
of the system,
 $\a\_{D}$ is the $D$-dimensional solid angle ($2\pi$ for $D=2$,
etc.). In fact, \cite{mil96} proves that this relation is not limited
to isothermal spheres, but holds for an arbitrary virialized system
subject to the above potential equation.

 This relation for deep-MOND, isothermal spheres ($D=3$, $A=4\pi G\ao$,
$\a_3=4\pi$):
\BE MG\ao={9\over 4}\vsm^2,  \label{relay} \EE
was first derived in \cite{mil84}; it
 was later generalized, first by \cite{gs92}
 to spherically symmetric systems with
general velocity distributions, and then, further, by \cite{mil94}
to arbitrary (not necessarily spherical) systems.
\par
Because the results of \cite{mil96} are not published yet
 we now give the proof of relation(\ref{dusna}),
for the (relatively simple)
 special case $D=2$, which constitutes a generalization of the
ELT result to arbitrary cylindrical systems.
 Following \cite{mil94} {\it mutatis mutandis}, we
consider a stationary, self-gravitating system, symmetric under $z$
translations, that is composed of
various particle species with masses $m_k$ and
distribution functions $\fk$, $\vr$, and $\vv$ being the position, and
velocity in the $x-y$ plane.
As usual, take the second time derivative of the quantity
\BE Q\equiv{1\over 2} \sum_k \int\dtr~\dtv~m_k~\fk~r^2, \label{qx} \EE
 which must vanish due to the
stationarity of the system (if the system is not stationary but still
bound its long-time average vanishes):
\BE \sum_k \int\dtr~\dtv~m_k~\fk~v^2+
\sum_k \int\dtr~\dtv~m_k~\fk~\vec r\cdot\vec a=0, \label{x} \EE
The first term in eq.(\ref{x}) is the
mass-weighted, 2-D, mean-square velocity, $\vsm$,
multiplied by $\mo$--the total line density
 of the cylinder.  In the second term we put $\vec a=-\gf$,
 where $\ff$ is the gravitational potential, to obtain
\BE \mo\vsm=
\sum_k \int\dtr~\dtv~m_k~\fk~\vr\cdot\gf. \label{xi} \EE
The $v$ integration, and the sum over species can now be performed
 to yield the standard result
\BE \mo\vsm=\int\dtr~\r(\vr)\vr\cdot\gf. \label{xii} \EE
Now, using the Poisson equation we can write
 eq.(\ref{xii}) as
\BE \mo\vsm=\ofpg\int \Delta\ff\vr\cdot\gf\dtr,   \label{ahs} \EE
which can be written as
\BE \mo\vsm=
 \ofpg\int \div[\vr\cdot\gf\gf-{1\over 2}(\gf)^2\vr]\dtr. \label{axx} \EE
Applying Gauss's theorem to write the integral as a surface integral
at infinity, and remembering that in the relevant geometry $\gf$
goes asymptotically as $\gf\rar-2G\mo\vr/r^2$, the integration gives
$G\mo^2$, and we finally get the general virial relation
\BE \vsm=G\mo.  \label{vr} \EE
When the system is not stationary, but
still bound, $\vsm$ in eq.(\ref{vr}) is replaced by its long-time
 average, $\overline{\vsm}$.
 \par
Note that, in fact, in deriving eq.(\ref{vr}) we do not make use of
the perfect cylindricity of the system. We only use
the Poisson equation, and the cylindrical
 boundary behavior of the potential at
 infinity. All still hold if the object looks like a uniform,
straight line at infinity, i.e., it can have nonuniformities of different
kinds: wiggles, modulations of the line density, etc., as long as it
is uniform on average. The problem is that in this case the system
cannot be expected to be stationary, so we cannot use the instantaneous,
observed value of $\vsm$, and we cannot easily relate it to
a line-of-sight velocity dispersion, as we do below.
\par
The 2-D root-mean-square
 velocity, $\vsm$, of an astronomical system
 cannot be measured from our vantage point,
and we would like to express it in terms of the
 one-dimensional, system-integrated, root-mean-square of the
 velocity component along
a line of sight perpendicular to the cylinder axis, $\sp$.
This can be done, for example,
 when $\sp$ is independent of the azimuthal viewing
angle, in which case
\BE \sp^2={1\over 2}\vsm, \label{vperp} \EE
and we may then write
\BE 2\sp^2=G\mo, \label{vira} \EE
which is the relation derived by ELT for their model filament.
Two instances in which eq.(\ref{vperp})
 is valid are: a. The velocity ellipsoid
is isotropic everywhere (e.g. in a gaseous system),
 but the system need be neither isothermal, nor
axisymmetric. b. The system is axisymmetric, but then the velocity
distribution need be neither isotropic, nor isothermal.
The ELT model is a special case of instance b.
\par
Another point to be made about the ELT model is that it can be
extended, at no further cost, to include a very dense core at the center
 of the filament. If we approximate this core by a line singularity
of finite line density $\m(0)$, then the structure equation of ELT is
modified only by replacing the anisotropy parameter $\b$ (see next
section or ELT for denotation)
 by $\hb\equiv\b+\emo$, where $\emo\equiv 2\m(0)G/\sr^2$.
The behavior of the density near the origin is now $\r\propto r^{-\hb}$.
convergence of the mass both near the origin, and at infinity is
assured for $\hb<2$, which sets an upper limit on the singular
 line density:
$\m(0)<(2-\b)\sr^2/2G$.
The relation between the ``gas'' line density (excluding the singularity)
and the "gas" velocity dispersion is now
\BE \m_t(gas)={2\sp^2\over G} -2\m(0). \label{lamde} \EE
and the total line density is
 \BE \m_{tot}={2\sp^2\over G} -\m(0). \label{lamhw} \EE
Our general relation(\ref{vira}) is seen to hold when we note that we
 must take in it the mass-weighted mean of $\sp^2$, including the
contribution of the line singularity. This latter must itself have
a $\sp^2(line)=\m(0)G/2$.
The line singularity can stand, for example, for an ELT model with
a radius scale much smaller than that of the ``envelope'' gas.
Such an extended model can help study possible departures from the
ELT model. For example, when the "gas" mass is negligible the singularity
 line density determines $\sp$ through $\m(0)=\sp^2/G$.

\section{Two idealized model filaments in MOND}
\par
 Obtaining a quantitative inkling of the MOND relation between
line density, radius, and velocity requires specific models.
 We now consider two such filament models.
Not enough is known about the velocity distribution, and radial-density
structure of real filaments to assess the degree of their adequacy.
Even if in their very outer parts actual filaments depart greatly from
these models, these may be applicable in the inner parts.

\par
 The first model makes the same assumptions as ELT's.
It consists of an infinitely long, axisymmetric, self-gravitating system
such that the azimuthal, and radial velocity dispersions--$\st$, and
 $\sr$ respectively--are position independent (accelerations along
  the axis are assumed negligible).
We assume that all accelerations are much smaller than $\ao$ so that
the deep MOND limit is taken.
The MOND Jeans equation can then be written as
\BE {\sr^4 \over \ao}\left(\deriv{ln~\r}{r}+{\b\over r}\right)
\left\vert \deriv{ln~\r}{r}+{\b\over r}\right\vert=-{\m(r)G\over r}.
 \label{jeans} \EE
Here, $\r(r)$ is the mass density at cylindrical radius $r$,
$\m(r)$ is the line density within radius $r$:
\BE \m(r)=2\pi\int_{0}^{r}\r(r')r'~dr',  \label{mu} \EE
 and
\BE \b\equiv 1-\st^2/\sr^2  \label{kauytr} \EE
is the anisotropy parameter.
 The system-integrated rms velocity component along
a line of sight perpendicular to the cylinder axis, $\sp$, is given by
\BE \sp^2=\sr^2(1-{\b \over 2}).  \label{sigp} \EE
Equation(\ref{jeans}) is derived along the lines detailed in
\cite{mil84}, and is based e.g. on the formulation of MOND in
\cite{bm84}.
\par
We shall see that the solution of eqs.(\ref{jeans})(\ref{mu}) leads to
a finite value of the asymptotic $\m$. Define then
$\mo\equiv \m(\infty)$, and use it to define length, and density scales
by
\BE \ro\equiv {\sr^4 \over \ao\mo G},~~~~\rhoo\equiv {\mo \over
2\pi\ro^2},  \label{scale} \EE
respectively.
 The deep-MOND Jeans equation can then be written as
\BE \z''-s\e^{-1}\z'+\e^{-1/2}\z^{1/2}\z'=0,  \label{eq} \EE
where $s\equiv 1-\b$, and
we use the dimensionless variables
\BE \e\equiv r/\ro,~~~~~\l\equiv \r/\rhoo, \label{dimle} \EE
and
\BE \z(\e)\equiv \m(\e\ro)/\mo=\int_{0}^{\e}\l(\e')\e'~d\e'.
 \label{aspo} \EE
Equation(\ref{eq}) is to be solved with the boundary conditions
$\z(0)=0$, and $\z(\infty)=1$.
\par
Near the origin, the third term in eq.(\ref{eq}) can be neglected
(it is the only term by which the Newtonian equation differs from
the MOND equation),
and we can solve the remaining equation to obtain there
\BE \z'(r\approx 0)\approx b\e^{s},~~~~\z(r\approx 0)\approx
{b\over s+1}\e^{s+1}.   \label{zero} \EE
Near the origin thus $\r(r)\propto r^{-\b}$, as in the Newtonian case.
Values $\b<0$ give nonmonotonic density laws, and we ignore these.
(The distribution function for such cylinders can be written as a
 function of the particle angular momentum, $J=r\st$, and the energy,
 $E=v^2/2+\ff(r)$, for unit mass particles, $f(\vv,r)\propto
J^{-\b}{\rm exp}(-E/2\sr^2)$, and is increasing with $J$ for $\b<0$.)
\par
At large $\e$ the second term is negligible, and we may put $\z\approx 1$
in the third term. Equation(\ref{eq}) than gives for the asymptotic
form of the density
 $\l(\e\rar\infty)\propto \e^{-1}{\rm exp}(-2\e\&{1/2})$, or
\BE \r\propto r^{-1} {\rm exp}[-2(r/\ro)\&{1/2}].  \label{asym} \EE
\par
 From eq.(\ref{zero}) we see that $\z$ is increasing near the origin.
By eq.(\ref{eq}) it cannot decrease at larger radii because if it has
a maximum $\z'$, and hence $\z''$ vanish there and $\z$ remains constant
from that point hence; it can also not increase indefinitely as can
be seen from eq.(\ref{eq}), and must then go to a constant at infinity,
has we have assumed all along.
\par
The solution for $\z$ can be found, numerically,
 by integrating eq.(\ref{eq}) out
from the origin, starting with expression(\ref{zero}),
 shooting to find the value of the parameter $b$ for which
 $\z(\infty)=1$.
We can avoid this trial-and-error
if we rescale the variables $\e$, and $\z$:
\BE \e=\c\eh,~~~~~\z=\o\zh,  \label{sac} \EE
such that a. our structure equation(\ref{eq}) remains of the same
form in $\eh$, and $\zh$, and b. near the origin $d\zh/d\eh=\eh^s$.
This can be shown to hold when $\o=\c^{-1}=b^{1/(2+s)}$.
The solution of eq.(\ref{eq}) for $\zh(\eh)$ is unique.
The asymptotic value of $\zh$ has to be identified with
 $\o^{-1}$, and can be used to find
\BE \z(\e)=\o\zh(\o\e).  \label{resac} \EE
(The equation can be solved analytically for one, non physical value of
 $\b=3/2$.)

We see then that $\ro$ appears as some radius scale in the density
 distribution of the isothermal cylinder. If it can be determined
for an actual filament together with $\sr$, we can
 determine $\mo$ through eq.(\ref{scale}) as
\BE \mo={\sr^4\over \ro\ao G},  \label{rmm} \EE
and, in terms of $\sp$ from eq.(\ref{sigp})
\BE \mo={4\sp^4\over(2-\b)^2 \ro\ao G},  \label{rmmt} \EE
\par
In order to use such relations to determine $\mo$ for real filaments we
 have to relate $\ro$ to some observable property such as the half-mass
radius, the projected
half mass radius, the radius, $\rh$,
 at which the surface density reaches a certain fraction of its central
 value, etc.. Take the last, for example, if we write
 the product $(2-\b)^2\ro$ appearing in the denominator of
 eq.(\ref{rmmt}) as $q\rh$ we have
\BE \mo\sim{4\sp^4\over q\rh\ao G},  \label{rmbs} \EE
 We find numerically that $q\approx 1$ for $\b=0$ (isotropic case),
$q\approx 1.5$ for $\b=1/2$, and $q\approx 2$ for $\b=1$ (radial orbits).

\par
The second model is contrived to afford analytic solution, but is not
less reasonable, in default of knowledge of real filaments.
It differs from the first in that the velocity dispersions are now
assumed to increase from the center out as the fourth root of the
 radius; thus
\BE \sr(r)=S^{1/4}r^{1/4},   \label{inca} \EE
and $\b$ is still constant (the projected, central velocity dispersion
does not vanish).
The constant $S$ can be used to construct a quantity
\BE \hm\equiv {S\over G\ao}, \label{hatmu} \EE
with the dimensions of a line density, which we use to define the
dimensionless line density
\BE \z(r)\equiv \m(r)/\hm. \label{zeta} \EE
The Jeans equation can then be written as
\BE \left[{1\over\r}\deriv{(r\&{1/2}\r)}{r}+{\b\over r\&{1/2}}\right]
\left\vert{1\over \r}
 \deriv{(r\&{1/2}\r)}{r}+{\b\over r\&{1/2}}\right\vert=-{\z(r)\over r}.
 \label{jeansi} \EE
This, in turn, can be written as
\BE \left(\deriv{ln~\r}{ln~r}+\d\right )
 \left\vert\deriv{ln~\r}{ln~r}+\d\right\vert =-\z(r), \label{hata} \EE
where $\d\equiv\b+1/2$, and
\BE \z(r)={2\pi\over\hm}\int_{0}^{r}\r(r')r'~dr'.  \label{aim} \EE
Near the origin, where $\z\approx 0$, we have
\BE \r(r\approx 0)\approx a r^{-\d}. \label{orig} \EE
Since, by definition, $\b\le 1$ the line density always converges at the
origin. However, for $\b>1/2$ the acceleration diverges there, and, in
the least, the deep-MOND assumption does not hold. These models
are thus only good for $\b\le 1/2$.

At large distances, where $\z\approx \mo/\hm$ ($\mo$, as before, is the
total line density)
\BE \r(r\rar \infty)\propto r^{-(\d+\t)}, \label{orsa} \EE
where $\t\equiv (\mo/\hm)^{1/2}$.
As the asymptotic form will be determined by the solution of
 eqs.(\ref{hata})(\ref{aim}), the value of $\t$ will be read off the
solution, and
 $\mo$ will be expressed in terms of $\hm$.
These equations possess a family of solutions whose members are
specified by the choice of the constants, $a$,
 dictating the behavior near
the origin. Defining
\BE \l(r)\equiv a^{-1}r^\d\r(r), \label{lama} \EE
and redefining the independent variable
\BE x =\left({r\over \rs}\right)^{2-\d},~~~~~\rs
\equiv\left[{8\pi a\over 9\hm(2-\d)^3}\right]^{{-1\over 2-\d}},
    \label{ouyt} \EE
we can write
\BE \deriv{ln~\l}{ln~x}=-{3\over 2}
\left[\int_{0}^{x}\l(x')~dx' \right]^{1/2}.
 \label{kada} \EE
This equation, with the boundary behavior $\l(x\approx 0)\approx 1$, has
 the unique solution
\BE \l(x)=(1+x^{1/2})^{-3}. \label{sola} \EE
Thus, the full density law is
\BE \r(r)=b \left({r\over \rs}\right)^{-\d}
\left[1+\left({r\over \rs}\right)^{(2-\d)/2}\right]^{-3},
  \label{solb} \EE
and
\BE \z(r)= {9\over 4}(2-\d)^2 \left[{(r/\rs)^{1-\d/2}\over
1+(r/\rs)^{(1-\d)/2}}\right]^{2}.    \label{soie} \EE
The radial scale length, $\rs$, characterizing the density law, is
arbitrary, and $b=a\rs^{-\d}$ is a
 function of $\rs$ through eq.(\ref{ouyt}).
In particular we read from eq.(\ref{soie}) that
\BE \mo={9\over 16}(3-2\b)^2{S\over \ao G}. \label{lpol} \EE
\par
As the case is with the first model
we can express $S$ in terms of different measures of the velocity
dispersion and scale length of observed filaments.
\par
 For example we can calculate the system average value of $\sr$, or $\sp$
\BE {2\over 2-\b}\sp^2=
 \langle\sr^2\rangle=S^{1/2}{\int\_{0}\&{\infty}r'^{3/2}\r(r')~dr'
\over \int\_{0}\&{\infty}r'\r(r')~dr'}.  \label{maie} \EE
The numerator is finite for $\b<1/2$, and we find
\BE {2\over 2-\b}\sp^2=
 \langle\sr^2\rangle=P S^{1/2}\rs^{1/2}, \label{hayu} \EE
where
\BE P={4(5-2\b)\over (3-2\b)^2}\left(
{\rm sin}{2\pi\over 3-2\b}\right)^{-1}.
\label{lths} \EE
The scale radius $\rs$ can be easily related, for example, to the
(space) half-mass radius, $r_h$, containing half the line density:
\BE \rs=(\sqrt{2}-1)^{2/(2-\d)}r_h. \label{klibs} \EE

Thus, in terms of $\sp$, and $r_h$
\BE \mo=Q{\langle\sp^2\rangle ^2 \over \ao G r_h}, \label{krts} \EE
where

\BE Q={9\over 4}(\sqrt{2}-1)^{-4\over 3-2\b}
\left({3-2\b\over 2-\b}\right)^2 P^{-2}.  \label{lnvc} \EE
 For $\b=-1/2$ $P=3/2$, and $Q\approx 6$.
 For $\b=0$ $P=40/9\sqrt{3}\approx 2.6$, and $Q\approx 2.5$.
 For $\b=.25$ $P\approx 4.9$, and $Q\approx 0.8$.
 For $\b$ approaching $1/2$ $Q$ goes to 0, but then most of the
contribution to the mean dispersion comes from larger and larger radii.

\section{A preliminary analysis of the Perseus-Pisces segment}
\par
Now we turn to the implications for the Perseus-Pisces super-filament.
 For reasons explained in their paper, ELT actually
 apply their method to only
a segment of the PP bridge of length of about $20h^{-1}~Mpc$, and diameter
of about $4h^{-1}~Mpc$  ($h=\ho/100~\kms Mpc^{-1}$).
 They find a system-integrated, line-of-sight
velocity dispersion of $\sp\sim430~\kms$,
estimate the B-band luminosity within their chosen section at
about $4\times10^{12}h^{-2}\lsun$.
 Using their $\s-\m$ relation they
get a value of $\mo\approx 8.5\times10^{13}\msun Mpc^{-1}$. All these
yield an estimated value of
 $\mlnl\sim 450h(\sp/430\kms)^2 \mlsunl$. We adopt all the above
system-parameter estimates. In addition,
we need a radius scale, $r$, for the filament, and choose to normalize
 $rh$  to $2~Mpc$, which seems the approximate typical radius from
 Fig. 1 of \cite{whg93}.
\par
We can estimate the MOND $M/L$ value in several ways
\bk
1. Estimate the acceleration $a$ and than correct the Newtonian ELT value
 by a factor $a/\ao$.
The value of $\ao$ as deduced by \cite{bbs91}, normalized properly to
 the assumed value of $\ho$ is $\ao\approx 2\times10^{-8}h^2\cmss$.
The acceleration at $r$ is
\BE a={\sr^2 \over r}\c,~~~~
\c\equiv\left\vert\deriv{ln~\r}{ln~r}+\b\right\vert,\label{acc}\EE
and so
\BE {a\over \ao}\approx 1.5\times10^{-2}\c
\left({\sr\over 430\kms}\right)^2
\left({rh\over 2Mpc}\right)^{-1}h^{-1}. \label{aao} \EE
Applying this correction factor to the ELT's Newtonian
 estimate we get the MOND estimate
\BE \mlm\sim 7\c
\left({\sr\over 430\kms}\right)^2
\left({\sp\over 430\kms}\right)^2
\left({rh\over 2Mpc}\right)^{-1}
\left({\ao\over2\times 10^{-8}h\&{2}\cmss}\right)^{-1}
\mlsun . \label{mlm} \EE
\bk
2. We can use the above isothermal-cylinder models with all due
 circumspection.
 For example, from eq.(\ref{rmbs}) we can write
\BE \mo\sim 2.6\times 10^{12}q^{-1}{\left(\sp\over 430\kms\right)^4}
\left({\rh\over 2~Mpc}\right)^{-1}
\left({\ao\over2\times 10^{-8}h\&{2}\cmss}\right)^{-1}\msun
 Mpc^{-1},   \label{rpps} \EE
where q is between 1 and 2 for $\b$ between 0 and 1.
With the B-band luminosity estimate of ELT
this gives
\BE \mlm\sim 13 q^{-1}
\left({\sp\over 430\kms}\right)^4
\left({\rh h\over 2Mpc}\right)^{-1}
\left({\ao\over2\times 10^{-8}h\&{2}\cmss}\right)^{-1}
\mlsun . \label{mlmsec} \EE
If we use eq.(\ref{krts}) from the second model we obtain an estimator
\BE \mo\sim 6\times 10^{11}Q{\left(\sp\over 430\kms\right)^4}
\left({r_h\over 2~Mpc}\right)^{-1}
\left({\ao\over2\times 10^{-8}h\&{2}\cmss}\right)^{-1}\msun
 Mpc^{-1},   \label{hondlk} \EE
and
\BE \mlm\sim 3 Q
\left({\sp\over 430\kms}\right)^4
\left({r_h h\over 2Mpc}\right)^{-1}
\left({\ao\over2\times 10^{-8}h\&{2}\cmss}\right)^{-1}
\mlsun . \label{mdimmm} \EE

\par
ELT discuss in detail many of the uncertainties that plague the analysis,
from doubtful model assumptions to difficulties in determining
system parameters. We here only stress further some of these and
add some comments peculiar to the MOND analysis.
\par
There are several factors that may cause an over-estimation of the
velocity dispersion $\sp$: 1. Contribution from variations in the
mean Hubble velocity along the filament. ELT show how this can be largely
removed by dividing the filament into segments and taking the
dispersion in each segment separately; but this was not done for the
actual Perseus-Pisces analysis. 2. Clumping along the filament that is
reckoned without can lead to a substantial over-estimate of $\sp$, as
ELT demonstrate on model filaments from N-body calculations. The PP
segment treated by ELT appears knotty to a degree
 on Fig. 1 of \cite{whg93}.
3. motion along the filament may contribute if this is not exactly
perpendicular to the line of sight everywhere.
In the case of the PP segment causes 1. and 3. seem to be of little
consequence (D. Eisenstein, private communication).
 Such uncertainties in $\sp$ are more crucial in MOND than in Newtonian
mass estimates, because the velocity enters the MOND mass
 estimate in the fourth power, not the second.
\par
Both the ELT and our method are global, and give the total $\m(\infty)$.
The luminosity, however, is taken by ELT only within a certain projected
radius (of about $2~Mpc$). The actual filament may well extend farther
in radius, and the estimated luminosity, uncertain as it is anyway, may
be systematically too low because of this, a fact which contributes
to an over-estimation of $M/L$.

\acknowledgements

I thank Avi Loeb for bringing the work of ELT to my attention prior
to publication, and for useful discussions, and Daniel Eisenstein
for comments on the manuscript.


\end{document}